\documentclass[5p,twocolumn,authoryear,times]{elsarticle}

\usepackage{lineno,hyperref}
\modulolinenumbers[5]
\hypersetup{colorlinks,citecolor=blue,linkcolor=blue,urlcolor=blue}

\journal{Astronomy \& Computing}

\usepackage{graphicx}	
\usepackage{amsmath}	
\usepackage{amssymb}	

\usepackage{natbib} 
\usepackage[usenames,dvipsnames]{color}
\usepackage{verbatim}
\usepackage{multirow}
\usepackage{booktabs}
\usepackage{subfigure}
\usepackage{colortbl}

\usepackage[T1]{fontenc}
\usepackage[utf8]{inputenc}
\usepackage[tracking=smallcaps]{microtype}
\SetTracking[no ligatures = f]{encoding = *,shape = sc}{0}

\usepackage{listings}
\usepackage[font=footnotesize,labelfont=bf]{caption}
\usepackage{footnote}
\makesavenoteenv{tabular}
\makesavenoteenv{table}
\usepackage{supertabular}
\usepackage{threeparttable}
\usepackage{longtable}
\usepackage{rotating}
\usepackage{lscape}

\usepackage{lineno}

\usepackage{bold-extra}
\usepackage{slantsc}
\usepackage{times}

\definecolor{verbgray}{gray}{0.9}
\definecolor{shadecolor}{rgb}{.9, .9, .9}
\definecolor{lightred}{rgb}{0.96,0.81,0.81}
\definecolor{verylightceleste}{rgb}{0.91, 0.93, 0.97}
\definecolor{lightceleste}{rgb}{0.75, 0.81, 0.98}
\definecolor{celeste}{rgb}{0.45, 0.58, 0.82}
\definecolor{white}{rgb}{1., 1., 1.}
\definecolor{darkblue}{rgb}{0.10, 0.35, 0.99}
\definecolor{darkblue2}{rgb}{0.10, 0.24, 0.59}
\definecolor{verydarkblue}{rgb}{0.05, 0.10, 0.24}
\definecolor{black}{rgb}{0., 0., 0.}
\definecolor{lightviolet}{rgb}{0.95, 0.89, 0.96}
\definecolor{violet}{rgb}{0.7, 0.28, 0.70}
\definecolor{darkviolet}{rgb}{0.42, 0.24, 0.41}
\definecolor{green}{rgb}{0.18, 0.41, 0.02}
\definecolor{lightgreen}{rgb}{0.62, 0.75, 0.43}
\definecolor{darkgreen}{rgb}{0.27, 0.34, 0.16}
\definecolor{verylightgreen}{rgb}{0.298039,0.952941,0.345098}
\definecolor{verylightgreen2}{rgb}{0.745098,1,0.654902}
\definecolor{yellow}{rgb}{1, 1, 0}
\definecolor{darkyellow2}{rgb}{1, 0.890196,0.427451}
\definecolor{darkyellow}{rgb}{0.87, 0.87, 0.03}
\definecolor{darkyellow3}{rgb}{1,0.933333,0.654902}
\definecolor{gray}{rgb}{0.14, 0.14, 0.15}
\definecolor{lightgray}{rgb}{0.70, 0.72, 0.74}
\definecolor{silver}{rgb}{0.91, 0.91, 0.91}
\definecolor{Blueb}{cmyk}{0.2,0,0,0}
\definecolor{lightorange}{rgb}{1,0.886275,0.619608}

\newcommand{\selavy}{\textsc{selavy}}
\newcommand{\aegean}{\textsc{aegean}}
\newcommand{\duchamp}{\textsc{duchamp}}
\newcommand{\caesar}{\textsc{caesar}}
\newcommand{\cutex}{\textsc{CuTEx}}
\newcommand{\sofia}{\textsc{SoFia}}
\newcommand{\asgard}{\textsc{asgard}}
\newcommand{\pybdsf}{\textsc{PyBDSF}}
\newcommand{\pyse}{\textsc{PySE}}

\newcommand{\hii}{H\textsc{ii}}

\usepackage{xcolor}
\definecolor{light-gray}{gray}{0.95}

\lstnewenvironment{codebkg}{%
  \lstset{
    escapechar=\&,
    basicstyle=\ttfamily\scriptsize,
    backgroundcolor=\color{light-gray},
    frame=single,
    breaklines=true,    
    framerule=0pt,
    columns=fullflexible
  }
 }
 {}

\begin{document}
\begin{frontmatter}

\title{Astronomical source finding services for the CIRASA visual analytic platform}

\author[1]{S. Riggi\corref{cor}}%
\ead{simone.riggi@inaf.it}
\author[1,2]{C. Bordiu}
\author[3]{F. Vitello}
\author[1]{G. Tudisco}
\author[1]{E. Sciacca}
\author[4,1]{D. Magro}
\author[5,1]{R. Sortino}%
\author[1,5]{C. Pino}
\author[6]{M. Molinaro}
\author[7]{M. Benedettini}
\author[8]{S. Leurini}
\author[1]{F. Bufano}
\author[1]{M. Raciti}
\author[1]{U. Becciani}

\cortext[cor]{Corresponding author}
\address[1]{INAF - Osservatorio Astrofisico di Catania, Via Santa Sofia 78, 95123 Catania, Italy}
\address[2]{Centro de Astrobiología (INTA-CSIC), Ctra. M-108, km. 4, 28850 Torrej\'on de Ardoz, Madrid, Spain}
\address[3]{INAF - Istituto. di Radioastronomia, Via Gobetti 101, 40127 Bologna, Italy}
\address[4]{Institute of Space Sciences and Astronomy, University of Malta, Msida MSD2080, Malta}
\address[5]{Department of Electrical, Electronic and Computer Engineering, University of Catania, Catania, Italy}
\address[6]{INAF - Osservatorio Astronomico di Trieste, Via G.B. Tiepolo 11, 34143 Trieste, Italy}
\address[7]{INAF - Istituto di Astrofisica e Planetologia Spaziali, Via del Fosso del Cavaliere 100, 00133 Roma, Italy}
\address[8]{INAF - Osservatorio Astronomico di Cagliari, Via della Scienza 5, 09047 Selargius (CA), Italy}






\begin{abstract}
Innovative developments in data processing, archiving, analysis, and visualization are nowadays unavoidable to deal with the data deluge expected in next-generation facilities for radio astronomy, such as the Square Kilometre Array (SKA) and its precursors. In this context, the integration of source extraction and analysis algorithms into data visualization tools could significantly improve and speed up the cataloguing process of large area surveys, boosting astronomer productivity and shortening publication time. To this aim, we are developing a visual analytic platform (CIRASA) for advanced source finding and classification, integrating state-of-the-art tools, such as the \caesar{} source finder, the ViaLactea Visual Analytic (VLVA) and Knowledge Base (VLKB). In this work, we present the project objectives and the platform architecture, focusing on the implemented source finding services.
\end{abstract}


\end{frontmatter}

\section{Introduction}
\label{sec:intro}
Innovative developments in data processing, archiving, analysis, and visualization are nowadays critical to deal with the data deluge expected in next-generation observing facilities for radio astronomy, such as the Square Kilometre Array (SKA) and its major precursors, i.e. the Australian Square Kilometre Array Pathfinder (ASKAP), MeerKAT, the Murchison Widefield Array (MWA) and the Low Frequency Array (LOFAR). The increased size and complexity of the archived image products will raise significant challenges in the source extraction and cataloguing stage, requiring more advanced algorithms to extract valuable scientific information in a mostly automated way.
Traditional data visualization performed on local or remote desktop viewers will be also severely challenged in the presence of very large data cubes, requiring more efficient rendering strategies, possibly decoupling visualization and computation, for example moving the latter to a distributed computing infrastructure.
The analysis capabilities offered by existing radio image viewers are currently limited to the computation of image/region statistical estimators or histogram displays and to data retrieval (images or source catalogues) from survey archives.
Advanced source analysis, from extraction to catalogue cross-matching and object classification, are unfortunately not supported as the graphical applications are not interfaced with source finder batch applications. On the other hand, source finding often requires visual inspection of the extracted catalogue, for example, to select particular sources, reject false detections, or identify the astronomical object class. Integration of source analysis into data visualization tools could therefore
significantly improve and speed up the cataloguing process of large surveys, supporting astronomers in the discovery of unknown and unexpected results,  boosting their productivity and shortening publication times. Interestingly, a recent survey \citep{Bordiu2020}, conducted among astronomers of different fields, has shown a surprising demand for visual analytics tools, denoted by $\sim$72\% of the respondents as one of the major needs in their research.\\
To tackle some of the highlighted challenges, we proposed to realize an integrated platform, dubbed CIRASA (Collaborative and Integrated platform for Radio Astronomical Source Analysis), for advanced source finding and classification driven by visual analytics techniques. CIRASA will integrate state-of-the-art tools, already in use within international collaborations, but also provide new developments to improve the source extraction and cataloguing capabilities (e.g. real vs spurious source identification, object classification, etc.) of existing finders and richer source visualization.
The platform is mainly tailored to the needs of the SKA and precursor radio community, aiming at providing a tool replicable at a larger scale in the SKA Regional Center infrastructure. Some of the provided features (e.g. source extraction and analysis algorithms) are, however, general purpose and may well serve the broader astronomical community in other wavelength domains (e.g. infrared, optical or gamma).\\This paper represents the first of a series of works, aiming at presenting the major components of the CIRASA project. It is organized as follows. In Section~\ref{sec:sci-context} we discuss in more detail the scientific context in which the CIRASA project was devised and is moving its first steps. In Section~\ref{sec:tech-context} we summarize the technological context of reference. In Section~\ref{sec:cirasa} we present the project, describing the current architecture and planned objectives. In Section~\ref{sec:caesar-srv} we present the source finding services, representing one of the major components that has been developed so far for the platform. Other components will be presented in follow-up papers. In Section~\ref{sec:deployment} we describe the current service deployment, reporting the reference testing metrics obtained on simulated images. Finally, in Section~\ref{sec:summary} we highlight the results achieved and the future activities.

\section{Scientific context}
\label{sec:sci-context}
SKA \citep{SKA1BaselineDesign,SKA1ConstructionProposal} will be the largest radio interferometer ever built, enabling sky surveys with unprecedented speed and level of detail ($\sim$nJy sensitivity, sub-arcsec spatial resolution, full frequency coverage from 50 MHz to 15 GHz), thus it is expected to revolutionize our knowledge of the Universe. Breakthrough discoveries are expected in several areas, from galaxy formation and evolution in the Epoch of Reionization to strong-field tests of gravity and the search for gravitational waves, but, possibly also in the Cradle of Life domain with the search for exoplanets and signals of extraterrestrial life. Significant discoveries are also expected in the study of our Galaxy. SKA will allow for a nearly complete census of radio-emitting Galactic objects, such as H\textsc{ii} regions, planetary nebulae (PNe) and supernova remnants (SNRs), currently prevented by the limited area and $uv$ coverage of past surveys.\\While SKA is currently starting the construction phase, its precursors have already completed the telescope commissioning phase and carried out scientific observations. ASKAP \citep{ASKAPSystemDesign}, for example, has currently completed the Early Science phase and first pilot survey observations \citep{ASKAP-PilotSurvey,ASKAP-RACSSurvey}, showing a great potential for serendipitous discoveries of new classes of objects and phenomena \citep{Norris2021}. The observations done in the Galactic plane \citep{Umana2021}, in particular, already achieved superior imaging performance compared to past surveys, enabling valuable scientific results to be obtained, even with an incomplete array \citep{Riggi2021}.\\MeerKAT had its first light in 2016 using 16 antennas, with the first science results published in April 2018 \citep{Camilo2018}. Data observations were carried out later on for all large science projects using 64 antennas. In the Galactic science context, preliminary scientific results from the MeerKAT Galactic Plane Survey (0.8-1.6 GHz) were recently reported \citep{ThompsonSKAConf2021,RiggiSKAConf2021}, and first data release is expected by the end of the year.\\At lower frequencies, MWA observations, started in mid-2013, are delivering hundreds of scientific works from the MWA collaboration (about 150 papers since 2015) or from external authors. The GaLactic and Extragalactic All-sky MWA (GLEAM) survey \citep{GLEAM1} has surveyed the sky south of declination +30$^{\circ}$ over a frequency range of 72–231 MHz.
Image and catalogue data covering a portion of the Galactic plane (|$b$|$\le$10$^{\circ}$; 345$^{\circ}$<$l$<67$^{\circ}$, 180$^{\circ}$<$l$<240$^{\circ}$) were recently released \citep{GLEAM2}. In this frequency range, LOFAR is also progressing similarly, releasing first data for both the Two-metre Sky Survey (LoTSS) \citep{Shimwell2019} at 120-168 MHz and the LBA Sky Survey (LoLSS) \citep{DeGasperin2021} at 42-66 MHz, and delivering hundreds of scientific works.

\section{Technological context and challenges}
\label{sec:tech-context}
Data processing and analysis challenges are, without any doubt, particularly relevant in SKA. Raw radio data produced from the antennas will be injected in the data processing pipeline at a rate of $\sim$TB/s and the amount of archived data, comprised of images, visibilities, and catalogues with millions of objects, is of the order of $\sim$EB/yr \citep{SKA1BaselineDesign,SKA1ConstructionProposal}. The volume and complexity of the final data products is so high that it will require more advanced analysis algorithms to extract the most important features in a mostly automated way, possibly exploiting data parallelism and emerging technologies in High Performance Computing (HPC) and Machine Learning (ML).\\
The analysis of SKA precursor observations is already raising significant challenges in the source extraction and cataloguing process at multiple levels, but also in data visualization, anticipating what will be needed to face with future SKA observations.
In the following section, we will briefly present the open issues and relative state-of-the-art for both topics, motivating the activities proposed for the CIRASA project, discussed in detail in Section~\ref{sec:cirasa}.
We will mostly consider the ASKAP Evolutionary Map of the Universe (EMU) survey \citep{Norris2011} (1.4 GHz, noise rms $\sim$10 $\mu$Jy/beam, angular resolution $\sim$10 arcsec, coverage $\sim$75\% full sky) as a reference case, taking in mind that similar challenges are present in all SKA precursors. Furthermore, we will focus on the analysis of 2D images only, thus not reviewing challenges and relevant tools specifically applicable to the analysis of 3D data cubes. These will be considered for future phases of project development.

\subsection{Source Finding and Classification}
The large field of view and the improved angular resolution of the SKA precursors have significantly
increased the typical size of the image data products, up to $\sim$16000$^{2}$ pixels per continuum survey tile in ASKAP \citep{ASKAP-PilotSurvey}, and $\sim$32000$^{2}$ pixels in SKA Data Challenge I (SDC1) simulations \citep{Bonaldi2021}. This introduced scalability issues in existing source finding algorithms, causing the processing time to exponentially increase, thus requiring the development of new finders able to distribute computing among multiple processing units. At present, none of the existing finders are able to fully exploit the potential offered by modern High Performance Computing (HPC) systems (based on multi-nodes and one or more accelerators per node) to scale up to very large images. Some finders, like \pyse{} \citep{Carbone2018}, \aegean{} \citep{Hancock2018} and \sofia\ \citep{SOFIA2}, have started to provide support for multithread runs, others also for multi-node processing, like \selavy{} \citep{WhitingHumphreys2012} and \caesar{} \citep{Riggi2016,Riggi2019}. Other finders \citep{Lucas2019} have invested in the optimization of existing algorithms, reaching optimal scalability performance on compact source extraction.\\
While the tools cited above were primarily designed for radio continuum observations (2D maps), other finders, like \sofia{} \citep{Serra2015} or \duchamp{} \citep{Whiting2012}, were specifically developed to tackle the even harder computational requirements of present spectral line observations (involving 3D position-position-velocity cubes).\\
The expected boost in sensitivity will allow for detecting millions of sources in large area surveys done with the SKA precursors, corresponding to an expected source density of a $\sim$1000s sources per deg$^{2}$. For example, the future EMU survey is expected to detect $\sim$70 million sources \citep{Norris2011}. At present, however, a much smaller density of few hundreds of catalogued sources per deg$^{2}$ is reported in ASKAP pilot surveys \citep{ASKAP-PilotSurvey,ASKAP-RACSSurvey}. Such a cataloguing process will require a level of automation and
knowledge extraction never reached before by state-of-the-art source finders. Although some finders used in the radio community have already been upgraded in this direction, many critical aspects still remain to be tackled, particularly for observations done in a dense and complex environment like the Galactic plane.

\subsubsection{Compact source extraction reliability}
The false detection rate (mainly due to over-deblending and image artefacts around bright sources) in many tested finders can indeed reach up to 20\% in fields with significant diffuse emission or extended sources \citep{Riggi2021}. A major effort is therefore needed to meet the high source reliability expectations (at least better than 99\%) of large area surveys. Spurious source rejection is, however, still manually performed in most of them. Besides being time-consuming and error-prone, this task is no longer feasible at the scale of SKA precursors. Moreover, the considerable efforts made in the visual source selection are typically not standardized in the adopted methodology and, unfortunately, often limited to the project under study without being re-used for the benefit of other projects. Although this stage cannot be completely avoided, particularly in the early project phase, investing time to develop improved quality selection criteria and
advanced rejection algorithms is of high priority. Promising results have been already obtained in this area with ML-classifiers (e.g. neural networks or decision trees) on simulated training datasets \citep{Riggi2019} and
on real datasets that were prepared by visual inspection of radio survey maps \citep{Mauch2003,Williams2019,Magro2021,Pino2021}.

\subsubsection{Automated detection of extended sources}
Several works attempted to quantify the completeness and reliability degradation (being reported around 10-20\%) of different source finders on both 2D images (e.g. \citealt{Hopkins2015}) and 3D data cubes (e.g. \citealt{Popping2012}) in presence of extended sources. All of these studies only tested performances on extended sources using the same algorithm developed for point-source extraction, considering one particular class of extended sources, generally modelled as elliptical gaussians with axes larger than the synthesized beam size. Extended structures with different morphologies and flux density profiles, such as the diffuse and faint objects found in the Galactic plane (e.g. large SNRs or \hii{} regions), are however mostly missed out by existing finders used in SKA precursor pipelines, highlighting a general lack of algorithms designed for this purpose.\\At present, only a few source finders \citep{Riggi2016,Robotham2018} provide dedicated algorithms for extended source extraction, but their performance, often measured on simulated data (e.g. see \citealt{Riggi2019}), is still well below to what is achieved for compact sources. These poor results force astronomers to eventually resort to a manual segmentation approach when extracting and delivering catalogues of extended sources \citep{Bordiu2021}.

\subsubsection{Source classification}
Source classification into known classes of objects is another poorly covered area to be already addressed in pilot survey observations.\\
In the Galactic plane, for instance, observations done with the SKA precursors \citep{Umana2021,Riggi2021} now enable the detection of almost all catalogued objects present in the surveyed field (\hii{} regions, SNRs, PNe, evolved stars), including a large fraction of sources previously considered as radio-quiet, thanks to the notable increase in sensitivity. Still, more than 90\% of the extracted sources have no counterparts at other wavelengths, or object identity information in existing astronomical databases. It is likely that the vast majority of unclassified objects are radio galaxies and \hii{} regions, while a smaller fraction is associated to PNe, evolved stars (Luminous Blue Variables, Wolf-Rayet) and SNRs. Completely new classes of objects are also to be expected. In this area, some progresses were recently reported \citep{Akras2019,Riggi2021} with classifiers based on traditional machine learning algorithms, making use of the infrared colours or the correlation between radio and infrared morphologies as the discriminant information among different classes of Galactic objects.\\Far from the Galactic plane, many activities are focused on the detection and classification of different flavours or morphologies of radio galaxies \citep{Wu2019,Clarke2019,Liu2019,Lukic2018,Lukic2019} or on the cross-identification of extragalactic radio sources and host galaxies \citep{Alger2018}, through deep convolutional neural networks. Many of these studies are carried out in the context of the Radio Galaxy Zoo (RGZ) project \citep{Banfield2015} and its ongoing follow-ups within some SKA precursors (for example ASKAP and LOFAR).

\subsection{Data visualization}
As we approach to the SKA era, two main challenges are to be faced in the data visualization domain: scalability and data knowledge extraction and presentation to users. The present capability of visualization software to
interactively manipulate input datasets will not be sufficient to handle the image data cubes expected in SKA ($\sim$200-300 TB at full spectral resolution). Even after frequency channel averaging, the requirement for next-generation data viewers is of the order of TB per cube and tens of GB per 2D image. Such a high data volume will require innovative visualization techniques and a change in the underlying software architecture models to decouple the computation part from the visualization. This is, for example, the approach followed by new-generation radio viewers such as CARTA\footnote{\url{https://cartavis.org/}}. CARTA uses a ``tiled rendering'' method and a client-server model, in which computation and data storage is performed on remote clusters with high performance storage, while visualization of processed products is performed on clients with modern web features, such as GPU-accelerated rendering.\\The volume and complexity of future SKA data will however require not only to import and visualize input data but also, mostly, to maximize the user perception efficiency, e.g. enabling for extraction of scientific results and discovery of new unexpected information from the processed data. 
To address these needs under a unified framework, visual analytics (VA) has recently emerged as the ``science of analytical reasoning
facilitated by interactive visual interfaces'' \citep{Yi2007}. VA aims to develop techniques and tools to support people in synthesizing information and deriving insight from massive, dynamic, unclear, and often conflicting data \citep{Keim2008}. To achieve this goal, VA integrates methodologies from information, geospatial and scientific analytics but also takes advantage of techniques developed in the fields of data management, knowledge representation and discovery, and statistical analytics. In this context, new developments have recently taken place in astronomy. As an example, the \emph{encube} framework \citep{Vohl2016} was developed to enable astronomers to interactively visualize, compare and query a subset of spectral cubes from survey data. The ViaLactea Visual Analytic application (VLVA) \citep{Vitello2018} (see Section~\ref{sec:cirasa}) allows for an integrated analysis of all new-generation surveys, combining the visualization of heterogeneous data, 2D intensity images, and 3D molecular spectral cubes.  

\begin{figure}
\centering%
\includegraphics[scale=0.42]{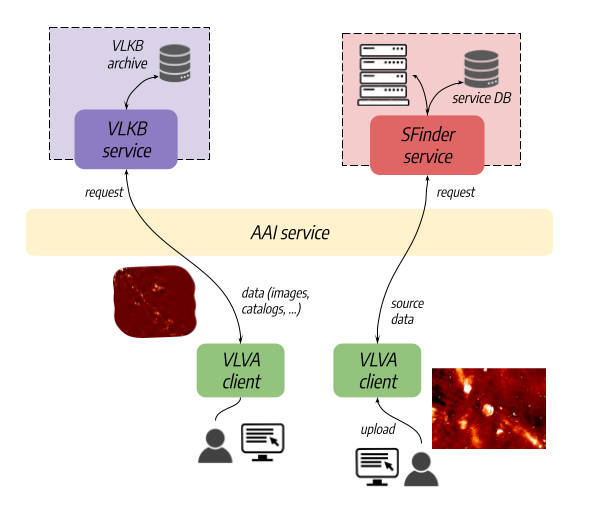}
\caption{High-level architecture of the CIRASA platform showing interactions among the major software components: ViaLactea Visual Analytic (VLVA) client, Authentication and Authorisation Infrastructure (AAI) service, ViaLactea Knowledge Base (VLKB) service and source finding (SFinder) service.}
\label{fig:cirasa-architecture}
\end{figure}

\section{The CIRASA project}
\label{sec:cirasa}
To address some of the highlighted challenges, we are developing a visual analytic platform, dubbed CIRASA. A high-level architecture diagram of the platform with major software components and expected data flow is shown in Figure~\ref{fig:cirasa-architecture}. The main components are a Visual Analytic client (VLVA) interfacing, through an authentication layer, with a series of services for source extraction, classification, and analysis, and a set of data collections (catalogues, images, cubes) exposing services for search, cutout and merge on top of the overall knowledge base archive (VLKB). All components are deployed in a distributed computing infrastructure.\\The platform, currently in development, should reach the following objectives:
\begin{enumerate}
\item Integrate existing compact and extended source finders (optimized for either continuum or spectral line images) into a common framework, exploiting each software's strengths and possibly combining their outputs to improve detection capabilities and source measurement (position, flux density, etc) accuracy;
\item Develop and integrate new source classifiers, exploiting innovative deep learning techniques, to enhance the performance of traditional source finders
and to enable creation of added-value catalogues;
\item Extend VLVA with interactive source visualization and validation functionalities, including both automated (e.g. by cross-matching with astronomical databases from the VLKB archive) and human-driven annotation functions to generate added-value catalogues and training data sets for classification scopes.
\end{enumerate}

\subsection{Software components}
A high-level overview of the CIRASA software components is reported as follows.

\subsubsection{Visual analytic client (VLVA)}
The ViaLactea Visual Analytic client (VLVA) \citep{Vitello2018} is a desktop interface implemented in C++ and based on Qt and VTK libraries. It represents the astronomer's entry point to platform resources. VLVA currently supports 2D and 3D visualization (e.g. through volume rendering, isocontours, slice views, etc) of images and data cubes, loaded either from the user local filesystem or from the remote VLKB archive upon valid authorization. The tool also enables the user to load and view catalogues of both compact and extended sources (currently only Galactic bubbles and filaments).\\VLVA is publicly available at {\footnotesize\url{https://github.com/NEANIAS-Space/ViaLacteaVisualAnalytics}} and distributed for both macOS and Linux (Debian/Ubuntu). More details are available in the online documentation at {\footnotesize \url{https://vlva.readthedocs.io/en/latest/index.html}}.\\New developments are occurring in different areas of the CIRASA project. Interfacing the VLVA client with source finding services described in this paper is one major area of extension. Further developments are planned to enhance the source visualization and analysis capabilities, following the use case described in section~\ref{subsec:objectives}. Such activities will be reported in a forthcoming paper.

\subsubsection{Knowledge Base archive and services (VLKB)}
The ViaLactea Knowledge Base (VLKB) \citep{Molinaro2016,Butora2019,Smareglia2019} is a large ($\sim$2 TB) archive of infrared, radio and molecular survey and source catalogue data ($\sim$40000 cubes and 2D images from $>$30 surveys), offering a series of service interfaces for catalogue access, dataset discovery, cutout creation (for 2D images as well as 3D cubes), and image/cube mosaicking through merging of adjacent areas of the sky stored in separate files.
A Table Access Protocol (TAP) interface \citep{Dowler2010} and a custom Multi-Order Coverage (MOC) based interface are furthermore available as defined by the International Virtual Observatory Alliance (IVOA), enabling the user to search and cross-match catalogues of both compact and extended objects, respectively.\\VLKB services are currently already interfaced with the VLVA client application. To support the main driving scientific use cases of the CIRASA project, the archive will also include the newest radio data produced in the Early Science phase of SKA precursors (ASKAP, MeerKAT) and simulated data generated in the SKA Science Data Challenges. New developments are also to be made in the VLKB service components to support cross-matching remote (e.g. stored in the VLKB archive) and local catalogues (e.g. residing in the local system of the VLVA client instance).

\subsubsection{Source finding services}
Source finding services, labelled as \emph{SFinder} in Fig.~\ref{fig:cirasa-architecture}, include one or more web applications, interfacing with various source extractor tools, enabling jobs to be launched on user data and outputs to be retrieved for visualization or further post-processing at the client level. In Section~\ref{sec:caesar-srv} we will present the architecture and implementation of one of these services, dubbed as \emph{caesar-rest}, currently integrating \caesar{} source finder and newer tools being developed as \caesar{} extensions or standalone applications. As discussed in Section~\ref{subsec:supported-apps}, we foresee that \emph{caesar-rest} can also integrate other source finders with limited efforts, so to provide a single \emph{SFinder} interface to other CIRASA services.


\subsection{Use cases}
\label{subsec:objectives}
A typical user workflow on the CIRASA platform, mainly arising from the experience gained with ASKAP \citep{Umana2021,Riggi2021} and MeerKAT \citep{Bordiu2021} early science data analysis, would therefore include the following major steps:

\begin{enumerate}
\item Load image or cube from local filesystem or from the VLKB archive using the image discovery, cutout, and mosaicking services;
\item Extract sources from image/cube using one or more integrated finders, according to the desired configuration, and draw them on the image/cube;
\item Apply group or filter operations to the source catalogue, e.g. select sources by position, region or name, apply selection criteria on source parameters or select/reject sources manually;
\item Inspect/analyse extracted sources individually (e.g. upon manual selection) or collectively (e.g. after a group operation) through dedicated
panels showing summary information, source parameter plots, analysis, or validation plots (e.g. source counts, sky distribution, etc);
\item Label or select sources in the following ways:
\begin{itemize}
  \item Cross-match source catalogue with desired astronomical catalogues through the VLKB and classify sources accordingly in an automated way;
  \item Apply pre-trained classifiers (e.g. for spurious source rejection or object classification) to the extracted source catalogue and relabel sources
accordingly in an automated way;
  \item Label/annotate sources interactively through the aided visual inspection tool;
\end{itemize}
\item Finalize the source catalogue and save outputs (tables in different formats, DS9 regions, etc).
\end{enumerate}

Many of the above functionalities are rather cross-domain and not only tied to the radio-astronomical community needs, making the platform reusable for astronomical data taken at different wavelengths.

\subsection{Long-term goals and synergies with other projects}
\label{subsec:project-synergies}
The CIRASA project fits well in the design activities of the SKA Regional Center (SRC) infrastructure, currently carried out within the SRC Working Groups (WGs), and in the supporting actions undertaken by many European countries for the setup of a network of supporting competence and computing centres. Indeed, one of the long-term goal of the project is making the platform available to SKA and precursor users, possibly deployed on SRC resources.\\Besides the technical design of the computing infrastructure, a major challenge is reducing the technological gaps for astronomers, promoting Open Science practices in research. A key role will be played in this context by the European Open Science Cloud (EOSC) programme, started by the EU Commission in 2015, and aiming to develop a trusted, virtual and federated environment, allowing researchers from different scientific disciplines to store, share, process and re-use research products following FAIR (Findable, Accessible, Interoperable, Reusable) principles. Under the EOSC initiative, the H2020 NEANIAS (Novel EOSC Services for Emerging Atmosphere, Underwater \& Space Challenges) project\footnote{\url{https://www.neanias.eu/}} is developing a Service Oriented Architecture (SOA) to deliver thematic services from different scientific communities into the EOSC. The first release of services tailored to the astrophysics and planetary science communities has recently been published, including tools for data management and visualization, for map making and mosaicking, and for automated structure detection \citep{Sciacca2021-adass}.\\
The CIRASA platform is adopting the same principles and technologies, reusing auxiliary services provided by the NEANIAS project (see Section~\ref{subsec:auxiliary-services}), and testing its services in the same deployment environment (see Section~\ref{sec:deployment}). Furthermore, all CIRASA service components, have been made publicly available (through Google or Microsoft account authentication) in the EOSC service marketplace\footnote{\url{https://marketplace.eosc-portal.eu/}}, although the currently available computing and storage resources do not allow supporting yet a large community of users, such as SKA Key Science Project (KSP) or SKA precursor survey teams. Nevertheless, the goal of both projects is to develop and deploy a system that can scale up once additional resources become available, either on the future EOSC cloud infrastructure, an SRC network node, or a smaller data centre (e.g. a Tier-3 cluster in a public research department, eventually part or not of the SRC network).

\begin{figure*}
\centering%
\includegraphics[scale=0.5]{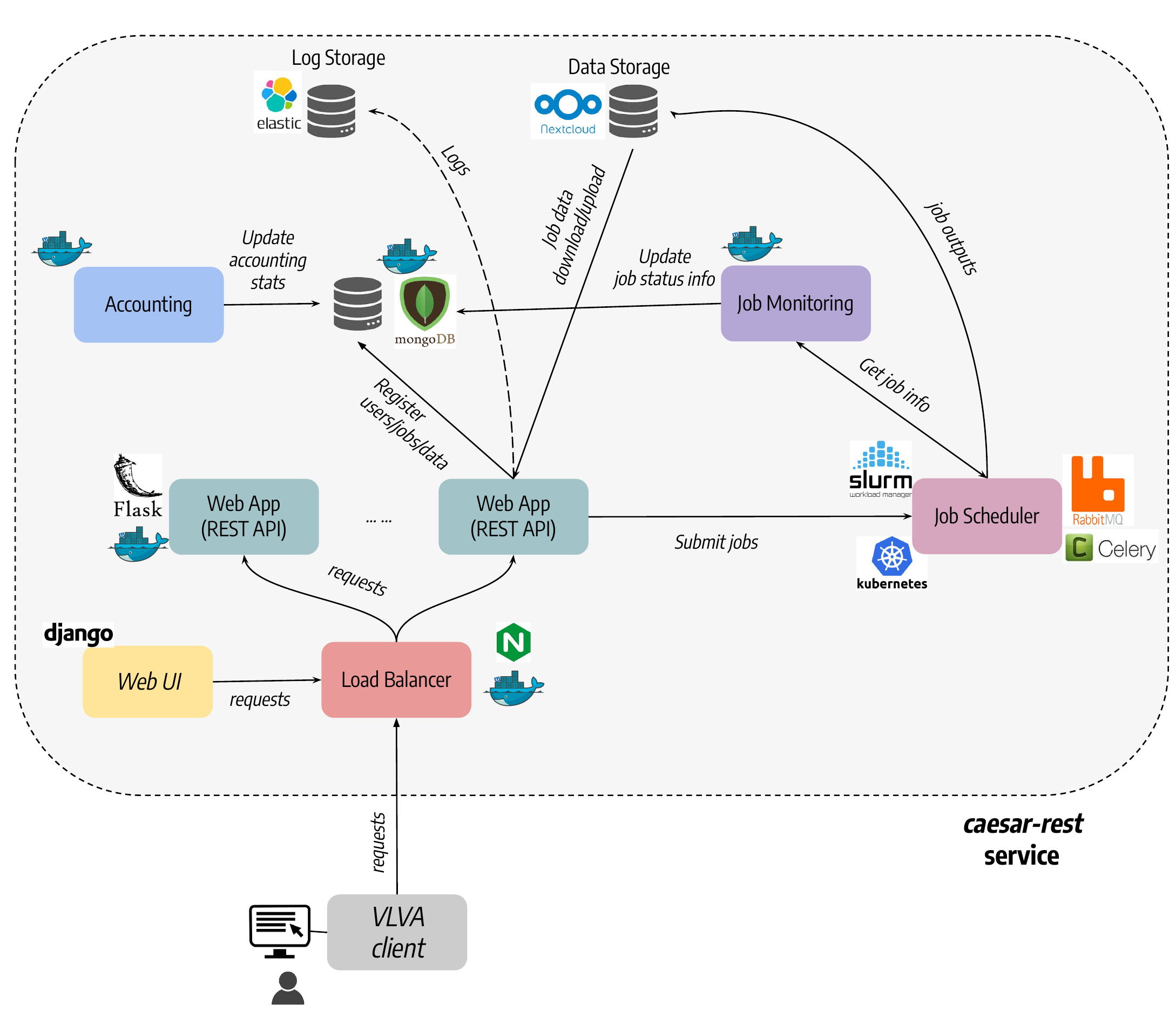}
\caption{Software components of the \emph{caesar-rest} service, representing the current implementation of CIRASA \emph{SFinder} service component, shown in a more schematic and abstract way in Fig.~\ref{fig:cirasa-architecture}. Other CIRASA components (e.g. the VLKB services) are not shown here.}
\label{fig:caesar-architecture}
\end{figure*}

\section{\emph{caesar-rest} source finding service}
\label{sec:caesar-srv}
For the CIRASA platform, we have developed a web service for source extraction and classification, named \emph{caesar-rest}. The software is developed in python and is publicly available at {\footnotesize \url{ http://github.com/SKA-INAF/caesar-rest}}, including API documentation, configuration options and instructions for service deployment.\\The architecture of the service consists of a few containerized microservices, shown in Fig.~\ref{fig:caesar-architecture}, deployable on a distributed computing infrastructure (see Section~\ref{sec:deployment}). The core service component is the web REST service, based on the Flask\footnote{\url{https://flask.palletsprojects.com/en/2.0.x/}} web framework and additional packages from the Flask ecosystem. In production, the Flask application is served by a uWSGI\footnote{\url{https://uwsgi-docs.readthedocs.io/en/latest/}} server, eventually replicated and run behind an NGINX load balancer. In ~\ref{appendix:config-options} and \ref{appendix:software-dependencies} we report a list of command-line configuration options and major software dependencies required by the REST service.\\A MongoDB\footnote{\url{https://www.mongodb.com}} database service is deployed to support the storage and retrieval of user data and job information (see details in Section~\ref{subsec:datamgt}).\\The job monitoring service supports periodical monitoring of user jobs and status info updates in the database. It is expressly required when using Kubernetes or Slurm job management (see Sections~\ref{subsubsec:kubernetes}. It is not required, instead, when using Celery (see Section~\ref{subsubsec:celery}), as, in that case, job monitoring is done by the deployed workers. Finally, the accounting service is not strictly mandatory, but, when deployed, computes some useful aggregated user data and job stats, making them available for querying (see Section~\ref{subsec:datamgt}) or displaying in the UI dashboard.\\The following paragraphs cover in more depth the service components and relative implementation.

\subsection{Data management}
\label{subsec:datamgt}
Service data (user images, job products) are stored in remote storage, directly accessible only by interested services (web applications and workers) through container volume mounting. Options tested were an NFS (Network File System) volume or a Nextcloud\footnote{\url{https://nextcloud.com}} storage managed with Rclone tool\footnote{\url{https://rclone.org/}} when deploying to OpenStack cloud instances or on a Kubernetes cluster.\\Both user data (e.g. location in storage, id, possible tags provided by the user, size) and job information (e.g. configuration options, id, status) are recorded in the MongoDB database following the naming conventions (\emph{dbname.username.jobs}, \emph{dbname.username.files}) for jobs and data collections, respectively. When the accounting service is deployed, an additional collection (\emph{dbname.username.accounting}) is populated with information about storage usage (for data and jobs) and job stats for each user as a function of time. Accounting information is periodically monitored (by default every 2 minutes) and aggregated over all users in a \emph{dbname.appstats} collection to provide some metrics at the application level.

\subsection{Job management}
\label{subsec:jobmgt}
User task submission requests are managed by the web application and pushed to a job scheduler queue for execution. Three different job schedulers and execution strategies are supported and can be configured at application startup: Kubernetes\footnote{\url{https://kubernetes.io/}}, Slurm\footnote{\url{https://slurm.schedmd.com/documentation.html}}, and Celery\footnote{\url{https://docs.celeryproject.org}}.\\A user job can assume the following possible state values during the run: 
{\footnotesize {\{\texttt{PENDING}, \texttt{RUNNING}, \texttt{SUCCESS}, \texttt{FAILURE}, \texttt{ABORTED}, \texttt{CANCELED}, \texttt{TIMED-OUT}}\}}. Values are self-explanatory, but not all of them can be mapped in the three architectures as discussed in the following paragraphs. Jobs are periodically monitored (by default every 30 seconds) by the job monitoring service and relevant information (e.g. state, status message, elapsed time, etc) is updated in the database.

\subsubsection{Celery}
\label{subsubsec:celery}
This is the most common approach encountered in Flask-based applications to handle long-running tasks. 
A Celery-based scheduling system requires a broker service to be added to receive task messages from the application and add them to a queue. Tasks of different source finding applications can be submitted to different queues. RabbitMQ\footnote{\url{https://www.rabbitmq.com}} and Redis\footnote{\url{https://redis.io}} are the two supported broker transports that can be selected and configured.\\One or more Celery workers must then be added to the system to consume the queued tasks. Workers executing a given source finding application, eventually customized in terms of consumable computing resources, subscribe and receive only the tasks queued for that application.\\Once received, tasks are processed by a Celery worker in the background and the process status is periodically monitored and updated in the MongoDB database. Celery allows for result backend components to be added to automatically store task status. In our application we have tested both Redis and MongoDB backends, using the latter as the default to have a unique database service in the architecture\footnote{In this case, Celery automatically creates an additional table (named \emph{celery\_taskmeta}) in the database to store task information}.\\This job management implementation has proven to work in our deployments, but we found these major limitations:
\begin{itemize}
\item Workers need to be constantly running, consuming the node resources allocated for them, even when no jobs are queued. This is not desired in a cloud infrastructure;
\item The architecture complexity is increased compared to the others described below, as two additional services need to be deployed (the broker and the task result backend, if different from the application database);
\item There is no straightforward way to allocate resources (CPU and memory) on demand. At present, the allocated resources can be configured on a per-worker basis.
\end{itemize}
These considerations motivated us to implement an additional work management schema.

\subsubsection{Kubernetes}
\label{subsubsec:kubernetes}
In this schema, jobs are submitted to a Kubernetes cluster that can eventually be the same hosting the application or an external one. 
Kubernetes has a default scheduler (\emph{kube-scheduler}) running in the control plane. When a Kubernetes job is submitted, the scheduler searches for a suitable worker node where to run job Pods\footnote{A Pod is the atomic deployment unit on a Kubernetes cluster, representing a single instance of a running process in it. A Pod contains a group of one or more application containers (such as Docker) that includes shared storage (volumes), a unique cluster IP address and information about how to run them. See \url{https://kubernetes.io/docs/concepts/workloads/pods/} for more details.}. A job is tracked and eventually restarted until termination conditions are met according to configurable specs (e.g. see \texttt{completionMode} and \texttt{backoffLimit} specs).
Finished jobs (either completed successfully or failed) and all dependant Pods can be cleaned up automatically from the system immediately or after a configurable time in seconds following the completion (see \texttt{ttlSecondsAfterFinished} spec). In practice, we found that this feature is not working properly in Kubernetes clusters with an older server version, for example, the one considered in Section~\ref{sec:deployment} (version 1.16.15). We therefore implemented a periodic clean-up of finished jobs in the job monitoring service. Jobs can be also deleted or suspended. This has the effect of cleaning up all dependant Pods permanently or until the job is resumed.\\
We made use of the Kubernetes Python client libraries\footnote{\url{https://github.com/kubernetes-client/python}} to manage jobs within our application. This requires to configure user authentication by passing the Kube cluster configuration and key/certificate files at the application startup. Once the client is initialized and configured, we mostly employed the BatchV1Api API \emph{create\_namespaced\_job}, \emph{read\_namespaced\_job}, \emph{delete\_namespaced\_job} functions to implement the job submission, monitoring, and clean-up logic. In this schema, a source finding job from one of the supported applications, is run by Kubernetes in a single Docker container Pod.\\Kubernetes API allows for a limited number of job states:
\begin{itemize}
\scriptsize%
\item \texttt{PENDING}: when the job is found in the job list, but its Pod is not active (e.g. running) nor failed or succeeded;
\item \texttt{RUNNING}: when the job Pod is reported as active;
\item \texttt{FAILED}: when the job Pod is reported as failed;
\item \texttt{SUCCESS}: when the job Pod is reported as succeeded;  
\end{itemize}
Another limitation is on the amount of job information reported, for example, the job elapsed time is only reported for successful jobs.

\subsubsection{Slurm}
\label{subsubsec:slurm}
In this schema, jobs are submitted to a Slurm cluster, typically external to the service. Slurm currently\footnote{Slurm version is v20.11 at the time of writing.} provides a REST API daemon named \emph{slurmrestd} enabling access to cluster resources upon valid authentication through RFC7519 JWT (JSON Web Tokens) tokens. Authentication can be configured in \emph{caesar-rest} service by passing the Slurm HS256-signed JWT user key at the application startup. This key is internally used by our Slurm client to initially generate the required JWT token (by default with 1 h duration), controlling its validity and regenerating it whenever needed.\\The same job management logic discussed in the previous section can be implemented around these API calls:
\begin{itemize}
\scriptsize%
\item POST \emph{/slurm/v0.0.36/job/submit}: for submitting a job, where the request body requires a job submission script and environment to be specified;
\item GET \emph{/slurmdb/v0.0.36/job/\{job\_id\}}: for retrieving the status of a job; \item DELETE \emph{/slurm/v0.0.36/job/\{job\_id\}}: for cancelling a job;
\end{itemize}
Jobs are submitted in this scenario using Singularity containers. Docker containers, used in the Kubernetes schema above, require root privileges to run and this is typically not granted for security reasons in co-shared Slurm clusters (e.g. department clusters, typically providing resources to multiple projects).\\As Slurm defines additional job states compared to our schema, we mapped them as follows: 
\begin{itemize}
\scriptsize%
\item \{\texttt{PENDING}, \texttt{SUSPENDED}\}$\rightarrow$\texttt{PENDING}
\item \texttt{RUNNING}$\rightarrow$\texttt{RUNNING}
\item \texttt{COMPLETED}$\rightarrow$\texttt{SUCCESS}
\item \texttt{CANCELLED}$\rightarrow$\texttt{CANCELED}
\item \{\texttt{FAILED},\texttt{NODE\_FAIL},\texttt{PREEMPTED},\texttt{BOOT\_FAIL},\texttt{DEADLINE}, \texttt{OUT\_OF\_MEMORY}\}$\rightarrow$\texttt{FAILURE}
\item \texttt{TIMEOUT}$\rightarrow$\texttt{TIMED-OUT}
\end{itemize}
Another difference with respect to Kubernetes is in the storage volume management. External data storage, e.g. the Nextcloud storage, is automatically mounted by the Kubernetes pods before actually executing the job. In this case, instead, they are mounted by the Slurm cluster administrator and Singularity job containers only need to bind to the defined mount point.

\subsection{Auxiliary services}
\label{subsec:auxiliary-services}

In the frame of the NEANIAS project, a layer of composite multi-tier services, integrated with the NEANIAS core infrastructure, was provided to support the open science lifecycle and the integration with the EOSC infrastructure. These include: an Authentication and Authorization Infrastructure (AAI), a Configuration Management Service, a Service Instance Registry, a Log Aggregator Service, Accounting and Notification services, and data depositing, sharing and exploration services. The auxiliary services currently exploited by the CIRASA platform are described in more detail below.

\subsubsection{Service authentication}
\label{subsec:aai}
User access verification on the service can be enabled at the application startup when in production mode. The only authentication protocol supported at present is Open ID Connect (OIDC)\footnote{\url{https://openid.net/connect/}}. Client requests without a valid auth token are rejected at this stage. In authorized requests, username information is extracted from the user email address field and used for all subsequent actions, e.g. to store data and job information in the database.

\subsubsection{Logging}
\label{subsec:logging}
The logging solution adopted is backed by an ELK stack, one of the most widely used stacks for collecting and processing application logs. The ELK stack is composed of three open source components, namely:

\begin{itemize}
\item Elasticsearch\footnote{\url{https://www.elastic.co/elasticsearch/}}, for storing and indexing application logs, making them searchable.
\item Logstash\footnote{\url{https://www.elastic.co/logstash/}}, for extraction and homogenization of log entries from different sources.
\item Kibana\footnote{\url{https://www.elastic.co/kibana}}, a visualization framework with aggregation and filtering capabilities.
\end{itemize}

On the application side, the service employs the Beats framework\footnote{\url{https://www.elastic.co/beats/}} for collecting the logs from the different architecture components and shipping them to the Logstash securely.\\The logging configuration can be configured at the application startup using a series of self-explanatory options (see~\ref{appendix:config-options}). The Filebeat service and logging to file must be enabled to collect application logs.

\subsection{Access layer}

\subsubsection{REST API}
\label{subsec:api}
\emph{caesar-rest} provides a REST API for:
\begin{itemize}
\item Uploading, downloading, or deleting input images from service data storage (see Section~\ref{subsec:datamgt});
\item Submitting source finding jobs using different supported applications to a workload management system, cancelling user jobs, or retrieving job status info and output data products (see Section~\ref{subsec:jobmgt});
\item Retrieving information about each supported source finder applications (e.g. configuration options usable for launching jobs);
\item Retrieving user accounting information of job and data minimal stats;
\end{itemize}
API specifications are reported in \ref{appendix:api}.

\subsubsection{Web interface}
\label{subsec:webui}
A web interface application has been developed in the context of the NEANIAS project to enhance accessibility and improve user experience for onboarding users. It provides authenticated, interactive access to the main service capabilities through a web browser, allowing for consuming the service REST APIs and the overall functionality in the first place, but also to guide as a valuable reference the development of a source finding interface for VLVA. 

The application is based on Django, a high-level Python framework for web development, and has been built following the standard MVT design pattern (Model-View-Template). The presentation layer follows the recommendations of the W3C on Cascading Style Sheets, Level 2 (CSS2), employing a customized version of the popular Twitter Bootstrap template, with a collapsible sidebar menu that provides access to the different features. User experience and interactiveness are boosted using multiple JavaScript libraries, such as jQuery.  REST APIs are consumed by means of Ajax requests, allowing for an asynchronous update of page contents and a smoother navigation.

\begin{figure*}
\centering%
\includegraphics[width=\textwidth]{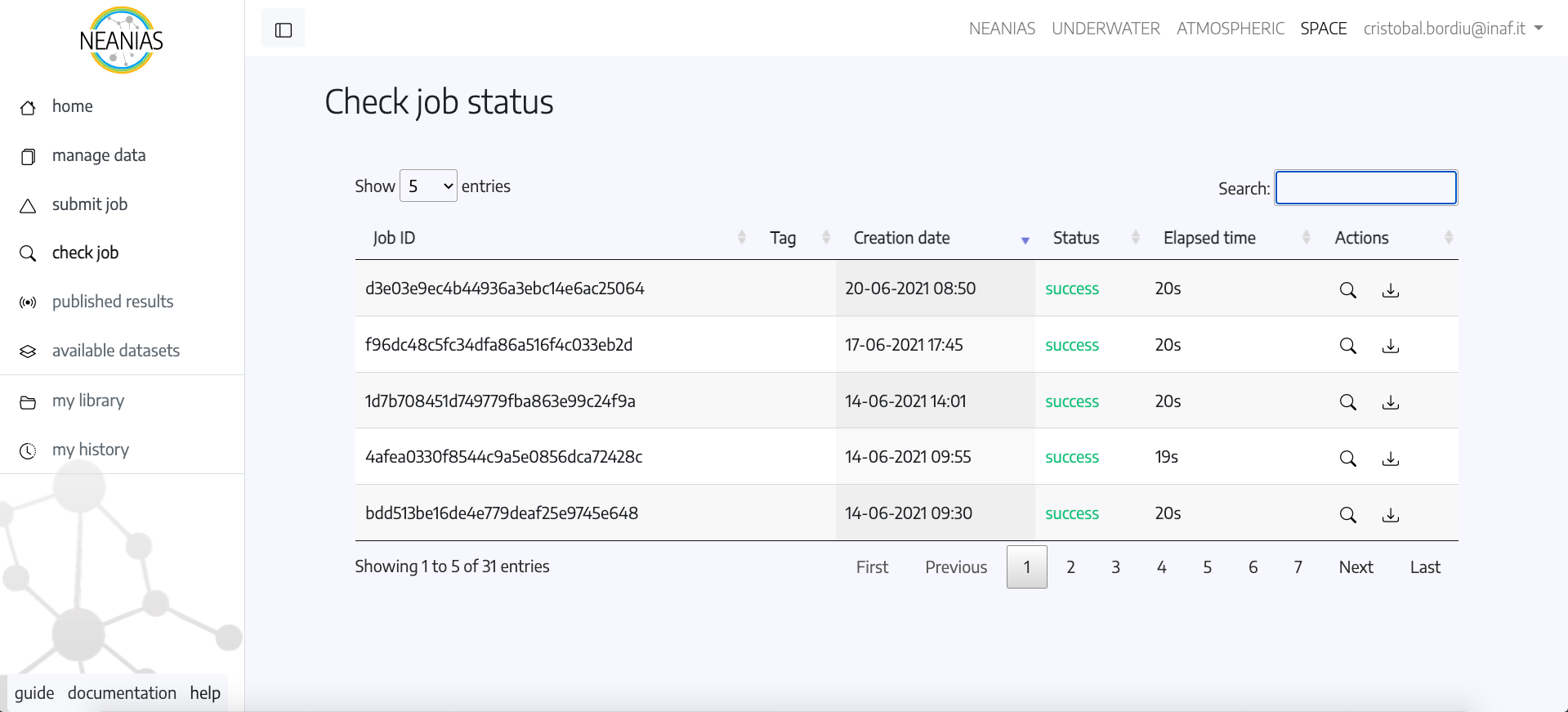}
\caption{List of submitted jobs in CAESAR UI.}
\label{fig:ui-joblist}
\end{figure*}

\begin{figure*}
\centering%
\includegraphics[width=\textwidth]{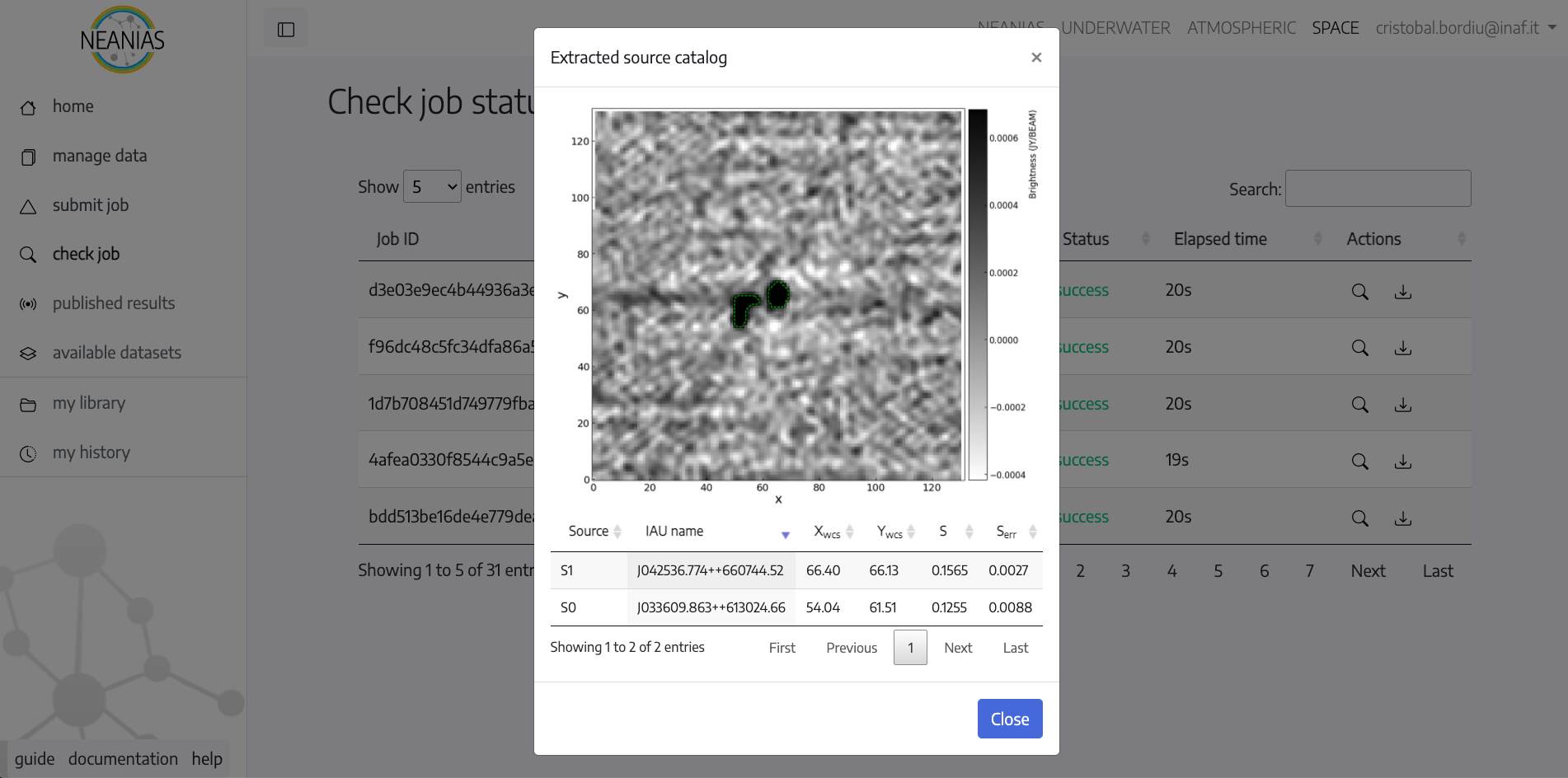}
\caption{Visualization of job results in CAESAR UI, displaying input image, with the extracted source contours overlaid, and an excerpt of the source catalogue.}
\label{fig:ui-preview}
\end{figure*}

The User-Centred Design methodology \citep{Cooper2014} has been loosely followed in the design of the user workflow, taking into account user requirements collected during the early stages of the NEANIAS project \citep{Sciacca2020}, and conducting several validation and feedback sessions with end users. The resulting workflow is intuitive: after logging in, the user is presented with a dashboard that compiles accounting information via simple graphs and widgets (e.g. number of jobs submitted, accumulated execution time, storage). Then, the user can:

\begin{itemize}
\item Manage files via the \textit{manage files} view, uploading images to be analysed --currently only FITS format supported--, and eventually adding convenient tags for easier identification. The files can also be downloaded back or removed from the system;
\item Submit source extraction jobs. The \textit{submit job} view provides a wizard that guides the user through the submission process: in the first step, the user selects which files to analyse; in the second, the user can customize the job, fine-tuning performance settings, source extraction settings, and background estimation settings; finally, in the last step, a summary of the selected options is displayed, and the user can assign a distinctive tag for the job. Multiple jobs can be submitted at a time if multiple files are selected, sharing the same settings and tag;
\item Preview and retrieve job outcomes. The \textit{check job} view presents a refreshable list of the submitted jobs, reporting the submission date, the current status, and contextual actions to cancel, preview, or download the job outputs (see Fig. \ref{fig:ui-joblist}). Currently, the application offers limited visualization capabilities, displaying a dismissible modal pop-up that shows the input image with extracted source contours along with an excerpt of the produced source catalogue (source name, position, and flux density), as displayed in Fig. \ref{fig:ui-preview}.
\end{itemize}

At the time of writing, the interface provides access exclusively to the \caesar\ source finder service. However, we note that the application is modular by design, facilitating further extensibility through the seamless addition of new components (e.g. access to new source finders or classifiers). In this regard, panels for performing runs with supported ML-based finders (see Section~\ref{subsubsec:ml-finders}) are planned to be added once their upgrade is complete.

\begin{figure*}
\centering%
\subtable[Response time]{\includegraphics[scale=0.4]{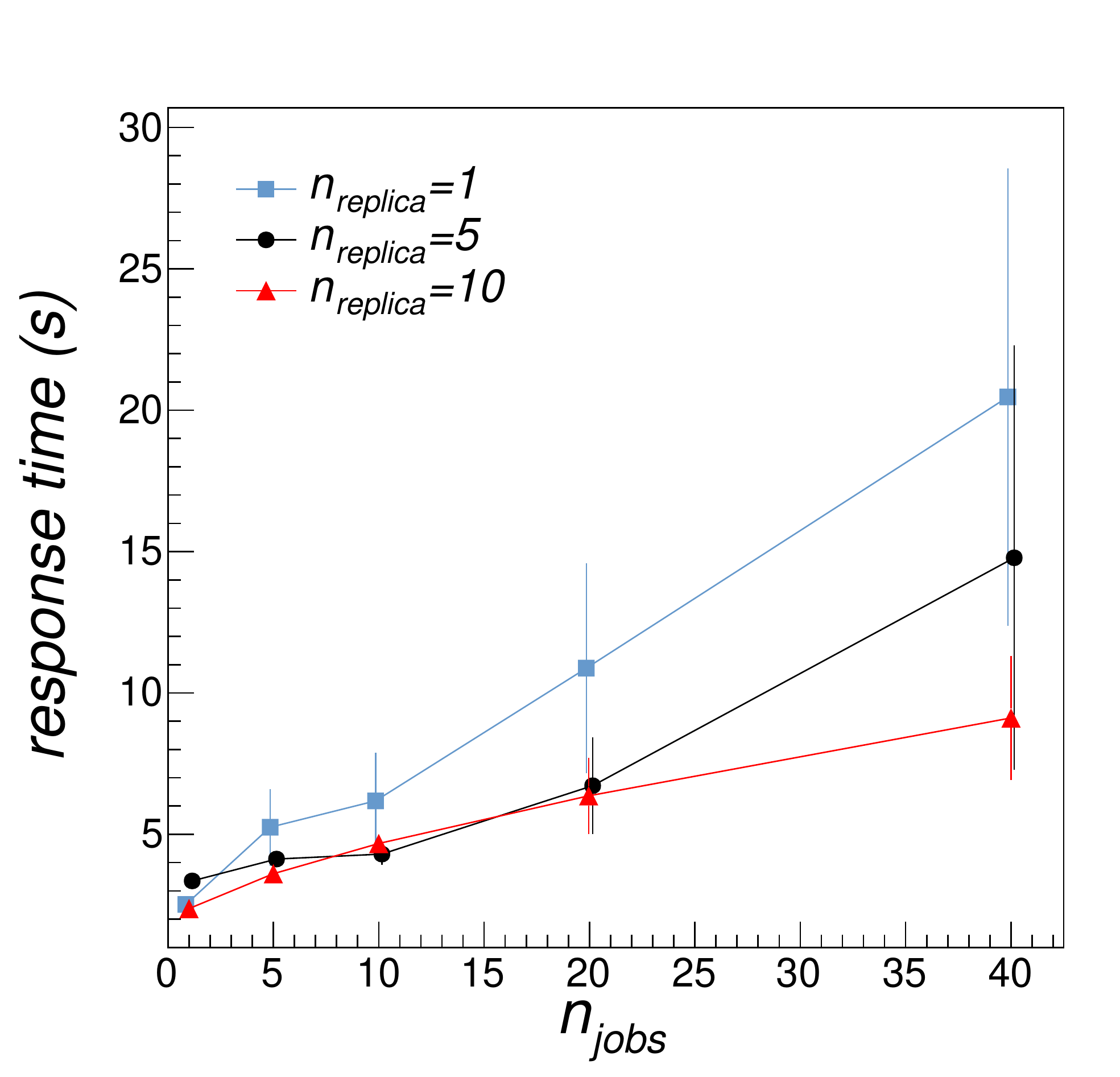}\label{fig:servicetests1}}%
\subtable[Job runtime]{\includegraphics[scale=0.42]{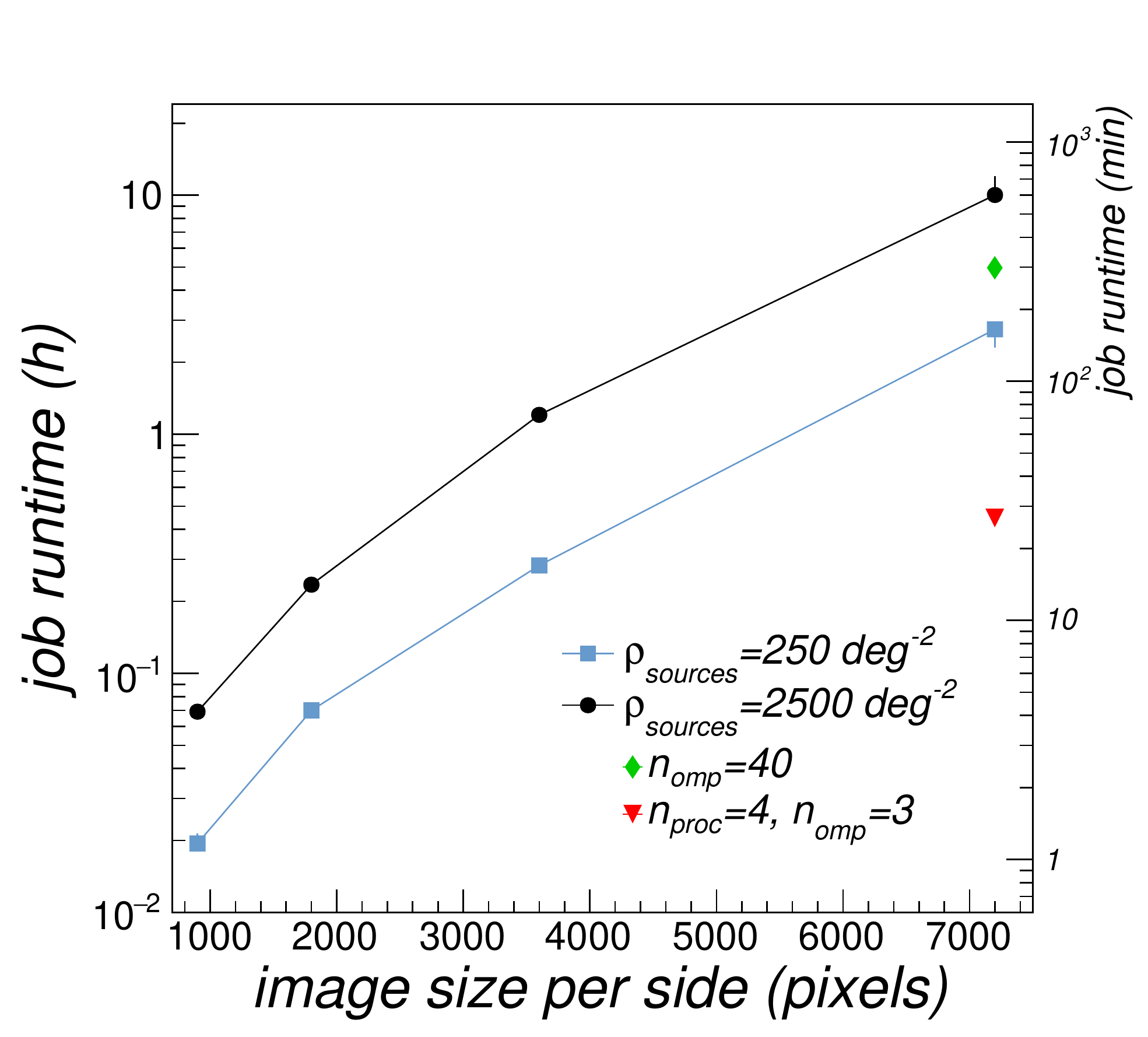}\label{fig:servicetests2}}%
\caption{Left: Median service response time taken when submitting a job to the \emph{caesar-rest} service as a function of the number of parallel job requests sent by a client. Results are reported for 1 (blue squares), 5 (black dots) and 10 (red triangles) web application replicas (see text). Right: Median runtime taken to complete \caesar{} jobs as a function of the input image size in pixels. Runtimes are reported for two different image source densities (see text): 250 deg$^{-2}$ (blue squares), 2500 deg$^{-2}$ (black dots). For images of size 7200$\times$7200 pixels and source densities of 2500 deg$^{-2}$, we report the runtimes obtained in multithread (n$_{threads}$=40, green diamond) and hybrid MPI parallel runs ($n_{proc}$=4 each with n$_{threads}$=3, red triangle).}
\label{fig:servicetests}
\end{figure*}

\subsection{Supported applications}
\label{subsec:supported-apps}
The service currently integrates the following applications: \{caesar,mrcnn,tiramisu\} (see next section). In the future, we plan to integrate other source finders widely used in the community for 2D images (e.g. \aegean{}, \pybdsf{}, \cutex{} and \emph{FilamentFinder} tools) or 3D cubes (e.g. \sofia{}) or new finders, for example based on deep learning models, either developed within the CIRASA project or within the radio community. In this respect, the integration of a new app only requires the provision of expected job options and application Docker containers. Integration in the CIRASA platform, requires, however, also a standardization of catalogue outputs across different finders, including content (e.g. the provided parameters) and format, currently ranging from custom tabular formats (CSV, ASCII) to other standards (e.g. JSON, VOTable). One possibility, currently under analysis in the project, foresees an extra processing step at the end of each source finder run, standardizing catalogue parameters and converting data into the desired format (likely JSON or a VO standard).

\subsubsection{CAESAR}
\caesar{} \citep{Riggi2016,Riggi2019} is a source finder for both compact and extended sources developed in the context of the ASKAP EMU survey. It currently supports batch parallel processing using two levels of parallelism (OpenMP and MPI) and provides both Docker and Singularity containers.\\It was recently employed to produce the compact source catalogue of the Scorpio field observed with ASKAP and ATCA \citep{Riggi2021}. Ongoing works \citep{Bordiu2021} are using it to produce compact and extended source catalogues from MeerKAT Galactic Plane survey data. Online documentation describing supported algorithms and configuration options is available at {\footnotesize{\url{https://caesar-doc.readthedocs.io/en/latest/}}}.

\subsubsection{ASGARD \& Tiramisu}
\label{subsubsec:ml-finders}
We recently developed two new tools for source detection and classification, dubbed \asgard{} (Automated Source, Galaxy, and Artefact R-CNN Detector) \citep{Magro2021} and Tiramisu \citep{Pino2021}, based on Mask-RCNN and U-Net deep learning models, respectively. They were trained on the same dataset, made of both public and private radio survey data (including ASKAP Early Science and pilot data), to detect three classes of objects: radio galaxies with extended morphology, compact sources and imaging sidelobes (artefacts). At the present stage, both tools are being upgraded to support processing on large images, rather than limited size cutouts. Another area of development aims to exploit both classifier predictions to boost performances of traditional finders, such as \caesar{}. Along this line, new state-of-the-art deep models and architectures are being tested on the existing training dataset to enhance current detection and classification performances, reported in the reference papers.

\begin{figure}
\centering%
\includegraphics[scale=5]{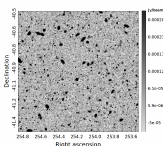}%
\vspace{-0.1cm}%
\caption{A sample simulated map (900$\times$900 pixels, 1 deg$^{2}$ area) produced for the testing campaign. FWHM synthesized beam is equal to 15" and Gaussian noise has a 20 $\mu$Jy/beam rms. Point sources were injected uniformly spaced and with flux density $S$ following $\exp^{-\lambda S}$ ($\lambda$=1.6) and ranging from 50 $\mu$Jy to 1 Jy. Extended sources were injected uniformly spaced from a 2D elliptical Gaussian model with randomized axes, up to a maximum major axis size of 3$\times$ beam size and minor/major axis size ratio varying from 0.4 to 1. Flux densities $S$ were generated according to $\exp^{-\lambda S}$ ($\lambda$=1.6) and ranging from 5 $\mu$Jy to 1 mJy. Total source density is equal to 2500 deg$^{-2}$, with a fraction of 10\% of extended sources.}
\label{fig:simmap}
\end{figure}

\section{Service deployment and testing}
\label{sec:deployment}
We have deployed the \emph{caesar-rest} service on different resource infrastructures, from single to multiple machines running on dedicated servers or on private clouds. Here we present the tests carried out with the service deployed on a Kubernetes cluster provisioned for the scopes of the NEANIAS project on the GARR OpenStack cloud\footnote{\url{https://cloud.garr.it/}}. Due to the limited resources available in this cluster, shared among other NEANIAS services, only the web application, database, and accounting/job monitoring services were deployed on the cloud. Job execution was instead performed on a Slurm cluster installed on a standalone server (Dell PowerEdge R740, 2$\times$Intel Xeon Gold 6248R 3.0 GHz, 48 cores, 512 GB memory) dedicated for the CIRASA project. The REST application was replicated on the Kubernetes cluster and put behind a load balancer service to improve the service scaling capabilities. Each replica requires 1 dedicated CPU and runs 2 uWSGI dual-threaded workers.\\
In Fig.~\ref{fig:servicetests} we report some metrics extracted from the performed tests. The left panel shows the median service response time (in seconds) obtained when submitting one or more jobs in parallel. Error bars are the median absolute deviations (MADs) for each test case. Tests were done with 1 (blue squares),  5 (black dots), and 10 (red triangles) REST application replicas running in the Kubernetes cluster. Response times are due to both the Flask and Slurm REST applications, currently deployed in different sites. As expected, the response times are increasing with the service load and overall improving as more replicas are available.\\For testing purposes, we produced several simulated radio maps of varying size and source density with both point and extended sources generated according to configurable parameters. Given the limited computing resources currently dedicated for the project, the maximum image size considered for the scalability tests was 7200$\times$7200, which is roughly comparable with image mosaic products of some SKA precursors surveys (e.g. LOFAR LoTSS, MeerKAT GPS, ASKAP Early Science surveys), but smaller by a factor 2 and 4.5 with respect to other surveys (e.g. the EMU Pilot and Rapid ASKAP Continuum Surveys) and SDC1 simulations, respectively. A sample test map is reported in Fig.~\ref{fig:simmap}.
In Fig.~\ref{fig:servicetests2} we report the median runtime of completed \caesar{} jobs for different simulated image sizes (in pixels) and source densities (in number of sources per deg$^{2}$). All runs were performed in serial mode, using a single core in the Slurm cluster. Computing times are particularly relevant for larger images and high source densities and are mostly due to the source fitting stage, as discussed in \cite{Riggi2019}. If sufficient resources are provided, however, \caesar{} jobs can be eventually run in parallel mode using OpenMP and MPI, reducing the computing times. For example, in Fig.~\ref{fig:servicetests2} we report the runtimes relative to the longest task (image size=7200$\times$7200 pixels, source density=2500 deg$^{-2}$) obtained in multithread (40 OpenMP threads, green diamond) and hybrid parallel (4 MPI processes, each with 3 OpenMP threads, red triangle) runs. As one can see, a modest speed-up is gained in the first case, while the computing time can be reduced by a factor of $\sim$20 by employing two levels of parallelism. In the first case, in fact, the input image is not partitioned into smaller sub-tiles (as in the second case, where serial tasks are operating on a smaller image) and some finder sub-tasks are known to have poor multithread scalability above 6-8 threads \citep{Riggi2019}.

\section{Conclusions and future work}
\label{sec:summary}
We have presented the CIRASA project, a visual analytics platform for astronomical source visualization and analysis, being developed mainly for meeting the radio astronomical community's needs in the SKA and its precursors era, but also usable with datasets at other wavelengths.
The platform consists of three main pillars: the VLVA client, the VLKB services, and the source finding services. The VLVA and VLKB services have already been integrated in the platform, and both are currently undergoing new developments to support the functionalities and requirements of the CIRASA and NEANIAS projects. New implementations will be described in forthcoming papers.\\
In this paper we described the architecture, implementation, and testing of the source finding service, also named \emph{caesar-rest} throughout the text. The service is currently deployed on a proto European Open Science Cloud infrastructure, backed up by dedicated CIRASA computing resources. This deployment was used to carry out performance tests on simulated radio maps to study the service response and scalability when varying the size of the input image, the radio source density, and the number of computing resources used for the application and the job submission. We have found that increasing the number of application replicas and the computing elements allow to significantly reduce the service response latencies, bringing the job runtimes to acceptable levels for a user even with large and densely populated maps.\\The service is currently undergoing a second major testing within a restricted community of astronomers (<50) selected in the NEANIAS project. User feedback will drive new developments to be made in the very near future, before moving to the integration with the VLVA.\\In the future we plan to integrate into the service other source finding applications widely in use in the radio community, additional source finding utilities provided with the \caesar{} tool (e.g. for source selection mainly), and new ML-based finders and classifiers being developed within the CIRASA project. The handling of the source catalogues produced by different finder algorithms is one of the functionalities that we foresee to develop both at the service and VLVA client level. This will ultimately provide the users with a wide selection of algorithms to be combined, leading to a considerable boost in source extraction performance for their analyses.

\section*{Acknowledgements}
Part of the research leading to these results has received funding from INAF under the PRIN TEC programme (CIRASA) and from the European Commissions Horizon 2020 research and innovation programme under the grant agreement No. 863448 (NEANIAS).

\appendix%
\onecolumn%

\section{\emph{caesar-rest} configuration options}
\label{appendix:config-options}
In Table~\ref{tab:config-options} we report a list of the command-line options currently available to configure the \emph{caesar-rest} service. The same options can also be configured for the provided Docker container (see \url{https://hub.docker.com/repository/docker/sriggi/caesar-rest}).

\begin{center}
\footnotesize%
\captionsetup{labelfont=bf}
\captionof{table}{
List of command-line options defined to configure the \emph{caesar-rest} service.}%
\label{tab:config-options}
\begin{threeparttable}
\begin{tabular}{lll}
\hline%
\hline%
Option & Default & Description \\%
\hline%
\rowcolor{lightgray}%
\multicolumn{3}{|c|}{MAIN OPTIONS}\\%
\hline%
\emph{-{}-datadir} & /opt/caesar-rest/data & Directory where to store uploaded data \\%
\emph{-{}-jobdir} & /opt/caesar-rest/jobs & Directory where to store jobs\\%
\emph{-{}-job\_scheduler} & celery & Job scheduler to be used. Options are: \{celery, kubernetes, slurm\}\\%
\hline%
\rowcolor{lightgray}%
\multicolumn{3}{|c|}{LOGGING OPTIONS}\\%
\hline%
\emph{-{}-loglevel} & INFO & Log level threshold. Options are: \{\scriptsize{DEBUG, INFO, WARN, ERROR}\}\\%
\emph{-{}-logtofile} & - & Enable log writing also to files\\%
\emph{-{}-logdir} & /opt/caesar-rest/logs & Directory to store log files\\%
\emph{-{}-logfile} & app\_logs.json & Log filename\\%
\emph{-{}-logfile\_maxsize} & 5 & Max file size in MB\\%
\hline%
\rowcolor{lightgray}%
\multicolumn{3}{|c|}{DB OPTIONS}\\%
\hline%
\emph{-{}-dbhost} & localhost & MongoDB database host\\%
\emph{-{}-dbname} & caesardb & Name of MongoDB database\\%
\emph{-{}-dbport} & 27017 & MongoDB database port\\%
\hline%
\rowcolor{lightgray}%
\multicolumn{3}{|c|}{AAI OPTIONS}\\%
\hline%
\emph{-{}-aai} & - & Enable service authentication\\%
\emph{-{}-secretfile} & - & File (.json) with client credentials for AAI service\\%
\hline%
\rowcolor{lightgray}%
\multicolumn{3}{|c|}{CELERY OPTIONS}\\%
\hline%
\emph{-{}-result\_backend\_host} & localhost & Celery result backend service host\\%
\emph{-{}-result\_backend\_port} & 27017 & Celery result backend service port\\%
\emph{-{}-result\_backend\_proto} & mongodb & Celery result backend service type. Options are: \{mongodb,redis\}\\%
\emph{-{}-result\_backend\_dbname} & caesardb & Celery result backend database name.\\%
\emph{-{}-broker\_host} & localhost & Celery broker service host.\\%
\emph{-{}-broker\_port} & 5672 & Celery broker service port.\\%
\emph{-{}-broker\_proto} & amqp & Celery broker service type. Options are: \{amqp,redis\}\\%
\emph{-{}-broker\_user} & guest & Celery broker service username\\%
\emph{-{}-broker\_pass} & guest & Celery broker service password\\%
\hline%
\rowcolor{lightgray}%
\multicolumn{3}{|c|}{KUBERNETES OPTIONS}\\%
\hline%
\emph{-{}-kube\_config} & - & Kubernetes cluster configuration file path \\%
\emph{-{}-kube\_cafile} & - & Path to Kubernetes client certificate authority file\\%
\emph{-{}-kube\_keyfile} & - & Path to Kubernetes client key file\\%
\emph{-{}-kube\_certfile} & - & Path to Kubernetes client certificate file\\%
\hline%
\rowcolor{lightgray}%
\multicolumn{3}{|c|}{SLURM OPTIONS}\\%
\hline%
\emph{-{}-slurm\_keyfile} & - & Path to Slurm rest service key file\\%
\emph{-{}-slurm\_user} & cirasa & Username enabled to run jobs in the Slurm cluster\\%
\emph{-{}-slurm\_host} & localhost & Slurm rest service host\\%
\emph{-{}-slurm\_port} & 6820 & Slurm rest service port\\%
\emph{-{}-slurm\_batch\_workdir} & - & Path to Slurm rest service key file\\%
\emph{-{}-slurm\_queue} & normal & Slurm cluster queue for submitting jobs\\%
\emph{-{}-slurm\_jobdir} & /mnt/storage/jobs & Path in which the job directory is mounted in Slurm cluster\\%
\emph{-{}-slurm\_datadir} & /mnt/storage/data & Path in which the data directory is mounted in Slurm cluster\\%
\emph{-{}-slurm\_max\_cores\_per\_job} & 4 & Maximum number of cores per node reserved for a job in the Slurm cluster\\%
\hline%
\rowcolor{lightgray}%
\multicolumn{3}{|c|}{VOLUME MOUNT OPTIONS}\\%
\hline%
\emph{-{}-mount\_rclone\_volume} & - & Enable mounting of Nextcloud volume through rclone\\%
\emph{-{}-mount\_volume\_path} & /mnt/storage & Mount volume path for container jobs\\%
\emph{-{}-rclone\_storage\_name} & - & rclone remote storage name\\%
\emph{-{}-rclone\_storage\_path} & . & rclone remote storage path to mount\\%
\hline%
\end{tabular}
\end{threeparttable}
\end{center}

\newpage%

\section{Software dependencies}
\label{appendix:software-dependencies}
In Table~\ref{tab:software-dependencies} we report a list of major software dependencies used in \emph{caesar-rest} service.

\begin{center}
\footnotesize%
\captionsetup{labelfont=bf}
\captionof{table}{
List of major software dependencies used in \emph{caesar-rest} service.}%
\label{tab:software-dependencies}
\begin{threeparttable}
\begin{tabular}{llll}
\hline%
\hline%
Software & Mandatory & Notes & References\\%
\hline%
\emph{Flask} & YES & - & \url{https://flask.palletsprojects.com/en/2.0.x/}\\%
\emph{uwsgi} & NO & Desired when running the service in production & \url{https://uwsgi-docs.readthedocs.io/en/latest/}\\%
\emph{flask-pymongo} & YES & - & \url{https://flask-pymongo.readthedocs.io/en/latest/}\\%
\emph{pymongo} & YES & - & \url{https://pymongo.readthedocs.io/en/stable/}\\%
\emph{flask\_oidc\_ex} & NO & Required when enabling service authentication & \url{https://pypi.org/project/flask-oidc-ex/}\\%
\emph{structlog} & YES & Needed to format log files to be sent to Logstash service & \url{https://www.structlog.org/en/stable/}\\%
\emph{celery} & NO & Required when enabling Celery job management & \url{https://docs.celeryproject.org/en/stable/}\\%
\emph{kubernetes} & NO & Required when enabling Kubernetes job management & \url{https://pypi.org/project/kubernetes/}\\
\hline%
\end{tabular}
\end{threeparttable}
\end{center}

\newpage%

\section{\emph{caesar-rest} APIs}
\label{appendix:api}
\emph{caesar-rest} APIs are described in the online software repository. Here we summarize the main functions.\\%
Users can upload their image data using the provided API method:

\begin{table}
\begin{center}
\footnotesize%
\caption{
List of job submission request data to be provided by user.
}%
\label{tab:jobsubmitdata}
\begin{tabular}{llll}
\hline%
\hline%
Field & Mandatory & Type & Description\\%
\hline%
\texttt{app} & YES & string & Job application name\\%
\texttt{tag} & NO & string & Assigned job label\\%
\texttt{data\_inputs} & YES & string & Input data uuid \\%
\texttt{job\_inputs} & YES & dictionary & Valid job options\\%
\hline%
\end{tabular}
\end{center}
\end{table}

\begin{codebkg}
POST caesar/api/v1.0/upload
\end{codebkg}

In case of success, the returned file \emph{uuid} has to be used for job submission or to download/delete the uploaded files, through these API methods, respectively:

\begin{codebkg}
GET /caesar/api/v1.0/download/{uuid}
POST /caesar/api/v1.0/delete/{uuid}
\end{codebkg}

where the following method allows for retrieving all the files uploaded by a user:

\begin{codebkg}
GET /caesar/api/v1.0/fileids
\end{codebkg}

To submit a job, clients need to use the following API method:

\begin{codebkg}
POST /caesar/api/v1.0/job
\end{codebkg}

where the expected request JSON data are described in Table~\ref{tab:jobsubmitdata}. 
Supported job options can be queried for each supported source finding application with the API method: 

\begin{codebkg}
GET /caesar/api/v1.0/app/{appname}/describe
\end{codebkg}

while a list of supported apps can be retrieved with this API method: 

\begin{codebkg}
GET /caesar/api/v1.0/apps
\end{codebkg}

Job submission returns the job identifier that has to be used to query the status of the job or cancel it using these API methods, respectively:
\begin{codebkg}
GET /caesar/api/v1.0/job/{job_id}/status
POST /caesar/api/v1.0/job/{job_id}/cancel
\end{codebkg}

To retrieve job outputs (in a zipped file format), the following method is provided:

\begin{codebkg}
GET /caesar/api/v1.0/job/{job_id}/output
\end{codebkg}

Some applications may also support additional methods for retrieving the individual job products. For example, \caesar{} supports retrieving the extracted source islands as ASCII table file or JSON format, using these API methods, respectively:

\begin{codebkg}
GET /caesar/api/v1.0/job/{job_id}/output-sources
GET /caesar/api/v1.0/job/{job_id}/sources
\end{codebkg}

Same functionality is available for fit component catalogue:

\begin{codebkg}
GET /caesar/api/v1.0/job/{job_id}/output-components
GET /caesar/api/v1.0/job/{job_id}/source-components
\end{codebkg}

A preview plot of extracted sources as PNG image file or Base64 encoded string can be obtained using these API methods, respectively:

\begin{codebkg}
GET /caesar/api/v1.0/job/{job_id}/output-plot
GET /caesar/api/v1.0/job/{job_id}/preview
\end{codebkg}

\end{document}